\documentclass[prl,twocolumn,showpacs,superscriptaddress,preprintnumbers]{revtex4-2}

\usepackage{mathrsfs,mathtools,bbold}
\usepackage{amsmath,amssymb}
\usepackage{verbatim}
\usepackage{graphicx}
\usepackage{ccaption}
\usepackage[dvipsnames]{xcolor}
\usepackage[colorlinks=true,linkcolor=blue,citecolor=BrickRed,urlcolor=MidnightBlue,filecolor=black]{hyperref}
\usepackage{tikz}
\usetikzlibrary{arrows.meta}

\DeclareFontFamily{OT1}{rsfs}{}
\DeclareFontShape{OT1}{rsfs}{m}{n}{ <-7> rsfs5 <7-10> rsfs7 <10->rsfs10}{} 
\DeclareMathAlphabet{\mycal}{OT1}{rsfs}{m}{n}

\newcommand{\e}{\epsilon}

\newcommand{\be}[1]{ \begin{equation}\label{#1} }
\newcommand{\ee}{\end{equation}}
\newcommand{\bea}[1]{\begin{eqnarray}\label{#1} }
\newcommand{\eea}{\end{eqnarray}}
\newcommand{\p}{\partial}

\newcommand{\refb}[1]{(\ref{#1})}

\renewcommand{\t}{\tau}

\begin{document}
 
\title{Carrollian Origins of Bjorken Flow}

\preprint{CPHT079.122022}

\author{Arjun Bagchi}
\email{abagchi@iitk.ac.in}
\affiliation{Indian Institute of Technology Kanpur, Kanpur 208016, India}
\affiliation{CPHT, CNRS, \'{E}cole polytechnique, Institut Polytechnique de Paris, 91120 Palaiseau, France}

\author{Kedar S. Kolekar}
\email{kedarsk@iitk.ac.in}
\affiliation{Indian Institute of Technology Kanpur, Kanpur 208016, India}

\author{Ashish Shukla}
\email{ashish.shukla@polytechnique.edu}
\affiliation{CPHT, CNRS, \'{E}cole polytechnique, Institut Polytechnique de Paris, 91120 Palaiseau, France}

\begin{abstract} 
Bjorken flow is among the simplest models of fluids moving near the speed of light ($c$) while Carroll symmetry arises as a contraction of Poincar\'{e} group when $c\to0$. We show that Bjorken flow and its phenomenological approximations are completely captured by Carrollian fluids. Carrollian symmetries arise on generic null surfaces and a fluid moving at $c$ is restricted to such a surface, thereby naturally inheriting the symmetries. Carrollian hydrodynamics is thus not exotic, but rather ubiquitous, and provides a concrete framework for fluids moving at or near the speed of light. 
\end{abstract}

%\pacs{11.25 -w, 11.25.Hf, 11.25Tq, 11.30 -j, 11.30.Ly}

\maketitle

\noindent {\em{\underline{Introduction}}}. Moments after the Big Bang, the universe existed in a dense hot soup of quarks and gluons moving at velocities very close to that of light, in a state that has come to be known as the Quark-Gluon Plasma (QGP). Experiments first at the Relativistic Heavy Ion Collider and later at the Large Hadron Collider at CERN have successfully recreated this QGP by smashing together heavy nuclei nearly at the speed of light. The theoretical understanding of this state of matter is at the heart of understanding the physics near the Big Bang. 

One of the main models used to describe high energy heavy-ion collisions has been the hydrodynamic description introduced by Bjorken \cite{Bjorken:1982qr}. This boost-invariant fluid flow, popularly known as Bjorken flow, depends only on proper time and not on rapidity, as we will go on to describe below. This phenomenological assumption holds true for the central rapidity region of the high energy collision. It is also one of the simplest models and has been very successful in describing this extreme regime where fluid velocity is near the velocity of light. We would be interested in understanding this old and very successful model in a completely new way in this paper. 

Our principal tool in this endeavour would be Carroll symmetry. 
The Carroll algebra arises in the peculiar speed of light $c\to0$ limit of the Poincar\'{e} algebra \cite{LevyLeblond, NDS}. This was long discarded as a mere mathematical curiosity without any physical applications whatsoever. Of late, however, there is a lot of interest in this degenerate relative of the Poincar\'{e} algebra, especially its conformal version, the conformal Carrollian algebra. It has been shown in \cite{Duval:2014uva}, following closely related observations in \cite{Bagchi:2010zz}, that this is isomorphic to the so-called Bondi-Metzner-Sachs (BMS) algebra which governs the asymptotic symmetries of Minkowski spacetimes at the null boundary \cite{Bondi:1962px, Sachs:1962zza}. Carrollian CFTs have hence emerged as a potential holographic dual of asymptotically flat spacetimes \cite{Bagchi:2010zz, Bagchi:2012cy, Barnich:2012aw, Bagchi:2012xr, Barnich:2012xq, Bagchi:2016bcd, Donnay:2022aba, Bagchi:2022emh, Donnay:2022wvx}. There are emerging links to condensed matter systems (e.g. the theory of flat bands \cite{Bagchi:2022eui}, fractons \cite{Bidussi:2021nmp}), cosmology and dark energy \cite{deBoer:2021jej}, and appearance of these symmetries in the context of tensionless string theory \cite{Bagchi:2013bga, Bagchi:2015nca,Bagchi:2020fpr}. Building on the connections to holography in flat spacetimes, and inspired by the original fluid/gravity correspondence \cite{Bhattacharyya:2007vjd}, a relation between Carrollian fluids on the null boundary with asymptotically Minkowski spacetimes in the bulk has been proposed \cite{Ciambelli:2018wre}. Carroll fluids have been further investigated in \cite{deBoer:2017ing, Ciambelli:2018xat, Petkou:2022bmz, Freidel:2022bai, Freidel:2022vjq}. But, perhaps unsurprisingly, given the nature of the Carrollian spacetimes where lightcones close down and everything becomes ultra-local, no applications of these exotic fluids have been found except for their aforementioned role in the flat version of the fluid/gravity correspondence 
{\footnote{We will discuss recent connections of Carroll hydrodynamics to the black hole membrane paradigm \cite{Donnay:2019jiz} near the end of the paper.}}.

In this paper, with the example of Bjorken flow, we show that contrary to views in the existing literature, Carrollian hydrodynamics is ubiquitous instead of being exotic, and arises whenever a fluid nears the velocity of light.

\smallskip

\noindent {\em{\underline{Central claim}}}. It is known, as we review below, that Carrollian symmetry governs generic null manifolds. Now, fluids that move at the speed of light are constrained to codimension one null hypersurfaces in the embedding spacetime. Our simple observation is that these {\em{fluids moving with the velocity of light will inherit the symmetries of the null hypersurface they are constrained to move on, and hence will be Carrollian}}. In fact, Carrollian symmetry, more concretely the systematic $c\to0$ expansion of relativistic hydrodynamics, would provide an excellent organizing principle for any fluid moving at speeds comparable to the speed of light. Carrollian hydrodynamics would emerge at the leading order and subsequent orders in $c$ would provide corrections to this leading approximation. In what follows, we show that Bjorken flow appears as a result of the leading Carrollian hydrodynamic formulation and the phenomenological assumptions that underly Bjorken flow are actually a consequence of Carroll symmetry. 

\smallskip

\noindent {\em{\underline{Bjorken flow}}}. Bjorken flow \cite{Bjorken:1982qr} describes spacetime evolution of dense and highly energetic state of matter produced in heavy-ion collisions as an ultrarelativistic fluid. One of the major simplifying assumptions of Bjorken flow is that all the interesting dynamics takes place along the beam axis \emph{i.e.} along the direction of collision of the two heavy nuclei. We will take this to be the $z$-axis. The flow assumes complete translation and rotation invariance in the transverse $x,y$ plane, leading it to become effectively two-dimensional. 

The second major assumption that underlies Bjorken flow is boost (or more precisely, rapidity) invariance, which is an assertion about the velocity profile of the fluid produced after the collision. Assuming that the collision occurred at time $t=0$ at $z=0$, the fluid exactly midway between the two receding nuclei at any later time $t$ continues to be at rest, whereas the fluid at the location $z$ moves with the longitudinal velocity $v = z/t$. In other words, at a particular instant in time $t$, the fluid at $z = 0$ is at rest, whereas the fluid at $z = \pm t$ is moving at the speed of light \footnote{When we speak of Bjorken flow, we will work in the natural units $c = \hbar = 1$. We will make the factors of $c$ explicit when we take the Carroll limit below.}.

It is useful to recast Bjorken flow in terms of the proper time-rapidity variables $(\tau, \rho)$, defined via
\begin{subequations}
\begin{align}
&\tau = \sqrt{t^2 - z^2},\label{eq:deftau}\\
&\rho = \tanh^{-1} v = \tanh^{-1}\left(\frac{z}{t}\right) = \frac{1}{2}\log\left(\frac{t+z}{t-z}\right).\label{eq:defrho}
\end{align}
\end{subequations} 
These expressions can be inverted to express $t, z$ in terms of $\tau, \rho$ as
\be{}
t = \tau \cosh \rho, \quad z = \tau \sinh \rho.
\ee
In terms of $\tau, \rho$ the Minkowski metric takes the form
\be{}
\label{eq:MilneMetric}
{\rm d} s^2 = - {\rm d} t^2 + {\rm d} z^2 + {\rm d} x_\perp^2 = - {\rm d}\tau^2 + \tau^2 {\rm d} \rho^2 + {\rm d} x_\perp^2,
\ee
where $x_\perp$ collectively denotes the transverse directions. This is known as the Milne metric. Note that the coordinate range $t \ge 0, |z| \le t$ corresponds to $\tau \ge 0, \rho \in \mathbb{R}$. Let us now look at the four-velocity $u^\mu$ of the fluid. In terms of the Lorentz parameter $\gamma = 1/\sqrt{1-v^2}$, it is given by $u^\mu = (\gamma,0,0, \gamma v)$. Using $v = z/t$, this becomes
\be{}
u^\mu = \frac{1}{\sqrt{t^2 - z^2}} \, (t,0,0,z) = (\cosh \rho,0,0, \sinh \rho)\, , \label{eq:four-velocity}
\ee
where $\rho$ is given by \eqref{eq:defrho}. Changing coordinates $(t,z) \to (\tau, \rho)$, we find the four-velocity becomes $u^\mu = (1,0,0,0)$. Thus, the assumption of boost invariance corresponds to the fluid being at rest in the $\tau, \rho$ coordinates.

Let us now consider the hydrodynamic equations for the fluid flow. Assuming that there are no additional conserved charges, the hydrodynamic equations are simply the equations for the conservation of various components of the energy-momentum (EM) tensor $T^{\mu\nu}$,
\be{}
\nabla_\mu T^{\mu\nu} = 0. \label{eq:hydroeqn}
\ee
We first consider the case of a perfect fluid, and will include viscous terms later in the paper. The EM tensor for a perfect fluid has the form: 
\be{}
T^{\mu\nu} = \left(\epsilon + P\right) u^\mu u^\nu + P g^{\mu\nu}\, , \label{eq:stress}
\ee
where $\epsilon, P$ are the energy density and pressure, which, in general, are functions of both $\tau, \rho$. In Bjorken flow, the assumption of boost/rapidity invariance leads to the velocity profile $u^\mu = (1,0)$ (suppressing the transverse directions), and implies that nothing depends upon the rapidity $\rho$. Thus $\epsilon = \epsilon (\tau), P = P(\tau)$, and so on. It is easy to see that the only non-zero components of the EM tensor \eqref{eq:stress} are $T^{\tau\tau} = \epsilon(\tau), T^{\rho\rho} = P(\tau)/\tau^2$ (which gives $T^{\tau}_{\tau} = -\epsilon$ and $T^{\rho}_{\rho} = P$). Also, the only non-vanishing Christoffel symbols for the metric \eqref{eq:MilneMetric} are $\Gamma^{\tau}_{\rho\rho} = \tau, \Gamma^\rho_{\rho\tau} = \Gamma^\rho_{\tau\rho} = 1/\tau$. Using these results in \eqref{eq:hydroeqn}, we find that Bjorken flow admits only one hydrodynamic equation:
\be{}
\frac{{\rm d} \epsilon}{{\rm d} \tau} = - \frac{\epsilon + P}{\tau}\, . \label{eq:BjorkenFlow}
\ee
Given the equation of state $P = P(\epsilon)$, \eqref{eq:BjorkenFlow} completely determines the time evolution of the system.

\smallskip

\noindent {\em{\underline{Carrollian symmetry}}}.
The Carroll group and its corresponding algebra arises in the vanishing speed of light limit of the Poincar\'{e} group in any number of spacetime dimensions $D$. The non-zero commutation relations for the Poincar\'{e} algebra are: 
\begin{equation}
\begin{split}\label{poin}
[M_{\mu\nu},M_{\rho\sigma}] &= 2\eta_{[\mu\sigma}M_{\nu]\rho} - (\rho \leftrightarrow \sigma), \\
[M_{\mu\nu},P_{\rho}] &= 2\eta_{\rho[\nu}P_{\mu]}, 
\end{split}
\end{equation}
where $\mu={0,1,2...,D-1}$, $P_{\mu}=-\partial_{\mu}$ are translation generators and $M_{\mu\nu}=x_{\mu}\partial_{\nu}-x_{\nu}\partial_{\mu}$ are Lorentz generators. Carrollian limit is achieved by taking the $c\to 0$ limit, or equivalently by taking the contraction $t\to \hat{\e} t$, $x^{i}\to x^{i}$, $\hat{\e}\to 0$. Under this limit, $M_{ij}$ stays the same, while
\begin{align}
     M_{0i} \to - \hat{\e} t\partial_{i} - \frac{1}{\hat{\e}}x_{i}\partial_{t}, \Rightarrow B_{i} \equiv \lim_{\hat{\e}\to 0}\hat{\epsilon} M_{0i} = - x_{i}\partial_{t}\, ,
\end{align}
where $B_i$ denotes the Carroll boosts. This gives us the Carroll algebra, with the non-zero commutators
\begin{align}
[M_{ij},M_{kl}]&=2\delta_{[il}M_{j]k} - (k\leftrightarrow l),\, [M_{ij},P_{k}] = 2\delta_{k[j}P_{i]},\nonumber\\
 [M_{ij},B_{k}]&= 2\delta_{k[j}B_{i]}, \, [P_{i},B_{j}]= \delta_{ij}H. \label{carr}
\end{align}
Notice that the Hamiltonian $H$ has now become a central element. 

\begin{figure}
\centering
\includegraphics[width=6.5cm]{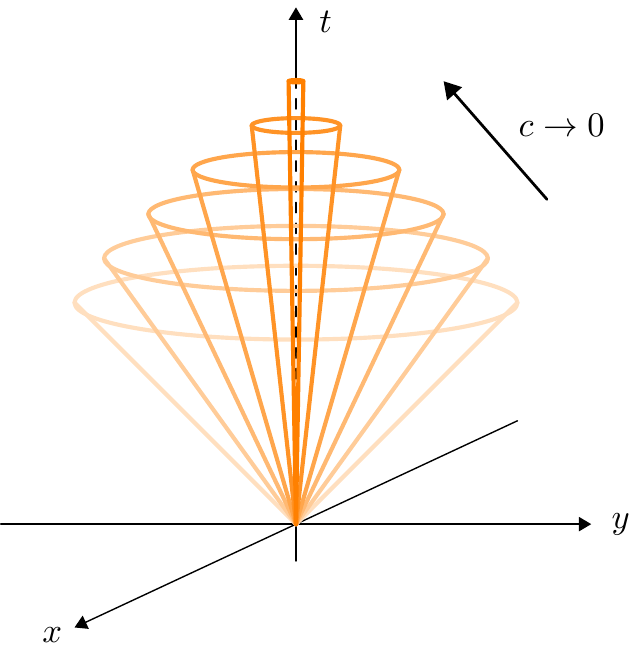}
\caption{Lightcones closing up in the Carroll limit.}
\label{fig1}
\end{figure}

Let us now attempt a more geometric understanding of the above. In the $c\to0$ limit, the (pseudo-)Riemannian nature of the spacetime manifold undergoes a drastic transformation, as the metric degenerates and lightcones close up (see fig.\ \ref{fig1}). In flat spacetimes, say in four dimensions, the covariant and contravariant metrics are
\be{}
\eta_{\mu\nu} = \begin{pmatrix} -c^2 & \mathbf{0}_{1 \times 3} \\  \mathbf{0}_{3 \times 1} &  \mathbf{1}_{3 \times 3} \end{pmatrix}, \quad 
\eta^{\mu\nu} = \begin{pmatrix} -\frac{1}{c^2} & \mathbf{0}_{1 \times 3} \\  \mathbf{0}_{3 \times 1} &  \mathbf{1}_{3 \times 3} \end{pmatrix}.
\ee
In the $c\to0$ limit, these degenerate and the covariant metric reduces to a spatial metric: 
\be{}
h_{\mu \nu} = \lim_{c\to0} \eta_{\mu\nu} = \begin{pmatrix} 0 & \mathbf{0}_{1 \times 3} \\  \mathbf{0}_{3 \times 1} &  \mathbf{1}_{3 \times 3} \end{pmatrix}. 
\ee
The contravariant metric becomes: 
\be{}
\Theta^{\mu \nu} = \lim_{c\to0} -c^2 \eta^{\mu\nu} = \begin{pmatrix} 1 & \mathbf{0}_{1 \times 3} \\  \mathbf{0}_{3 \times 1} &  \mathbf{0}_{3 \times 3}  \end{pmatrix} = \theta^\mu \theta^\nu, 
\ee
where $\theta^\mu$ is the vector $(1,0,0,0)$. It is obvious that in this case, we also have
\be{}
h_{\mu\nu} \theta^{\mu} =0.
\ee
It is now easy to generalise the above to what is called a Carrollian manifold $\mathcal{C}$. A $D$-dimensional Carrollian manifold $\mathcal{C}$ is defined by the pair $(h_{\mu\nu}, \theta^{\mu})$, where $h_{\mu\nu}$ is a tensor field of rank $(D-1)$ and signature $(0, + \ldots +)$ and $\theta$ is a vector field that generates the kernel of $h$ \cite{Henneaux:1979vn,Duval:2014uoa}. One can further define a connection on the manifold $\mathcal{C}$, which we will not specify here. The Lie algebra of the vector fields $\xi = \xi^\mu\p_\mu$ arising out of the isometry conditions
\be{}
\mathcal{L}_\xi h_{\mu\nu}=0= \mathcal{L}_\xi \theta^{\mu}
\ee
gives rise to the Carroll Lie algebra. Solving the above equations for flat Carroll structures described above, one gets
\be{}
\xi^0 = a + f(x^i), \quad \xi^i = \omega^i_{\, j} x^j + g^i.
\ee
This algebra is infinite dimensional, due to the undetermined function $f(x^i)$. By requiring a connection compatible with the pair $(h_{\mu\nu}, \theta^{\mu})$, the function $f(x^i)$ can be made linear. The resulting algebra of $\xi^\mu$ then reduces to the finite algebra \refb{carr} we obtained by a contraction. 

\smallskip

\noindent {\em{\underline{Carroll hydrodynamics}}}.
Carrollian fluids are fluids that flow on Carrollian manifolds. We will work with a particular coordinate chart on a Carroll manifold $\mathcal{C}$ given by $(t,\textbf{x})$, where the degenerate metric and its kernel take the form
\begin{equation}\label{deg_met}
    \text{d}\ell^2=a_{ij}(t,\textbf{x})\text{d}x^{i}\text{d}x^{j}, \quad \text{e}_{\hat{t}}=\frac{1}{\Omega}\partial_t.
\end{equation}
Here $\Omega=\Omega(t,\textbf{x})$, and the kernel has as the dual form 
\begin{equation}
    \vartheta^{\hat{t}}=\Omega\text{d}t-b_i\text{d}x^i,
\end{equation}
where $b_i(t, \textbf{x})$ is the Ehresmann connection on $\mathcal{C}$. As discussed, Carrollian geometries can be obtained as $c\to0$ for a (pseudo-)Riemannian geometry. It is particularly convenient to take this limit in the Papapetrou-Randers (PR) parametrisation
\begin{equation}\label{R-P_met}
    \text{d}s^2 =  - c^2\left(\vartheta^{\hat{t}}\right)^2+a_{ij}(t,\textbf{x})\text{d}x^{i}\text{d}x^{j}\,.
\end{equation}
The metric above reduces to the geometric structure \eqref{deg_met} as $c\to0$.

We now derive Carrollian fluid equations in the $c\rightarrow 0$ limit of relativistic hydrodynamics \cite{Petkou:2022bmz}. Consider a perfect fluid in the PR coordinates
\begin{equation}
    T^{\mu\nu}=(\epsilon+P)\frac{u^{\mu}u^{\nu}}{c^2}+P g^{\mu\nu},
\label{eq:stress_c}
\end{equation}
where compared to \eqref{eq:stress}, we have now restored the factors of $c$. A useful parametrization for the fluid velocity $u^\mu \partial_\mu = \gamma \partial_t + \gamma v^i \partial_i$, satisfying normalization $u^{\mu}u_\mu=-c^2$, is 
\begin{equation}
\label{eq:gen_para}
\gamma = \frac{1+c^2 \boldsymbol{\beta \cdot b}}{\Omega \sqrt{1- c^2 \beta^2}}\, , \quad v^i = \frac{c^2 \Omega \beta^i}{1+c^2 \boldsymbol{\beta \cdot b}}\, ,
\end{equation} 
with $\beta^i(t,\bf{x})$ being a Carrollian vector field parametrizing the fluid velocity. In the limit $c \rightarrow 0$ we get
\begin{equation}
\gamma = \frac{1}{\Omega} + \mathcal{O}(c^2)\, , \quad v^i = c^2 \Omega \beta^i + \mathcal{O}(c^4)\, .
\end{equation}
A natural ansatz for the scaling of energy density and pressure as $c\rightarrow 0$ is \cite{Ciambelli:2018wre}
\begin{equation}
\label{eq:Carroll_Limit_1}
\epsilon = \varepsilon+\mathcal{O}(c^2), \quad P = p + \mathcal{O}(c^2).
\end{equation}
The components of the EM tensor \eqref{eq:stress_c} are hence
\begin{equation}
    T_{00}=\varepsilon\Omega^2+\mathcal{O}(c^2),\, T^{ij} = pa^{ij}+\mathcal{O}(c^2),\, T^{0i} = \mathcal{O}(c).
\end{equation}
Using \eqref{eq:hydroeqn}, we find that up to $\mathcal{O}(c^2)$ terms the Carroll fluid equations are
\begin{subequations}\label{EnergyP}
\begin{align}
    \frac{1}{\Omega}\partial_t\varepsilon &= -\theta\left(\varepsilon+p\right),\\
\hat{\partial}_i p &= - \varphi_i\left(\varepsilon+p\right) - \left(\frac{1}{\Omega} \partial_t + \theta\right) \left(\varepsilon + p\right) \beta_i.
\end{align}
\end{subequations}
Above we use
\begin{equation}
\hat{\partial}_i \equiv  \left( \partial_i + \frac{b_i}{\Omega} \partial_t \right)\!, \theta \equiv \frac{1}{\Omega}\partial_t\ln{\sqrt{a}},  \varphi_i \equiv \frac{1}{\Omega}\left(\partial_tb_i+\partial_i\Omega\right)\!,
\end{equation}
where $\theta, \varphi_i$ are the Carrollian expansion and acceleration, respectively. Also, $a \equiv {\rm det} \, a_{ij}$.

\smallskip

\noindent {\em{\underline{Emergence of Bjorken flow from Carroll hydrodynamics}}}.
To connect our discussion of Carrollian hydrodynamics to Bjorken flow, we now adopt a particular choice of the PR gauge: 
\be{milneCarroll}
\Omega=1, \,\,\,\, b_i=-\beta_i, \,\,\,\, a_{ij}\text{d}x^i\text{d}x^j = \tau^2\text{d}\rho^2 + \text{d}x^2+\text{d}y^2.
\ee
Let us motivate our choice here. Recall the original Milne metric \refb{eq:MilneMetric}. When we consider Bjorken flow, we wish to be at small proper time and at high rapidity. This essentially means taking $\t \to \hat{\e} \t$ along with $\rho \to \rho/\hat{\e}$, with $\hat{\e}\to0$. In the Bjorken picture, the dimensionless parameter $\hat{\e}$ controls the boost invariance approximation. On the other hand, $\hat{\e}$ plays the role of the speed of light in the Carroll limit. Taking this limit on the Milne metric gives us the PR parametrisation \refb{milneCarroll}. 

\begin{figure}[t]
\centering
\includegraphics[width=8.6cm]{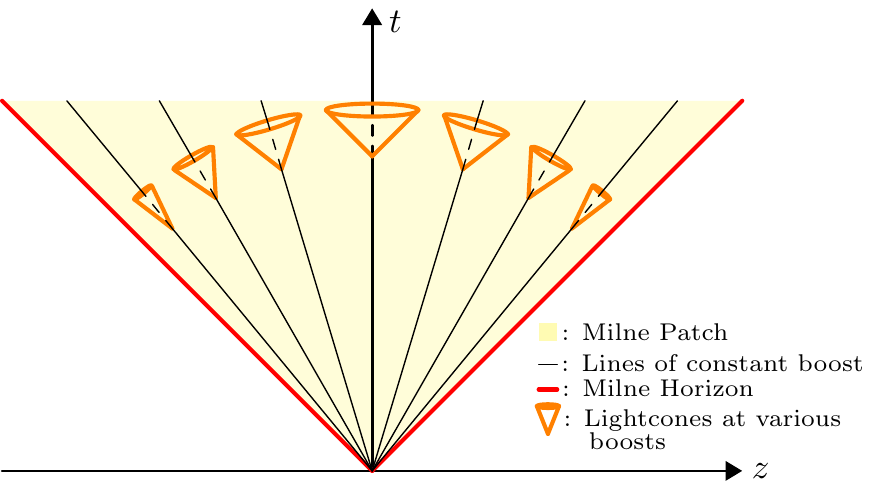}
\caption{Collapsing lightcones with increasing boosts in the Milne patch.}
\label{fig2}
\end{figure}

In coordinates adapted to Milne, \eqref{milneCarroll}, the Carroll fluid equations \eqref{EnergyP} give
\begin{equation}
    \partial_\tau\varepsilon=-\frac{\varepsilon+p}{\tau}, \quad \partial_i p = 0. 
\end{equation}
The first equation is the Bjorken equation, \refb{eq:BjorkenFlow}, courtesy of \eqref{eq:Carroll_Limit_1}. The second equation says that the pressure depends only on $\tau$, and in particular is independent of $\rho$, the rapidity. This is the phenomenological approximation of Bjorken. We have thus {\em derived Bjorken's equation and his phenomenological approximation of independence from rapidity} by considering Carrollian hydrodynamics adapted to Milne coordinates {\footnote{The attentive reader may point out that we have not shown that $\varepsilon = \varepsilon (\t)$, as required by Bjorken's approximation. This is implicit. The Carroll fluid will have an equation of state  $p=p(\varepsilon)$. We have shown $p=p(\t)$. Hence $\varepsilon=\varepsilon(\t)$.}}.

Fig.\ \ref{fig2} provides an intuitive picture of the process. In Milne coordinates, as we boost the fluid and move to higher and higher rapidities, lightcones around the fluid close down. In the limit of infinite boost, the fluid reaches the speed of light and now lies along the Milne horizon, which is a null surface. Although not visible in fig.\ \ref{fig2}, the lightcones in this limit have completely closed down, and Carrollian symmetries arise on the horizon, which is a null surface. The fluid naturally imbibes these symmetries and hence is described by Carrollian hydrodynamics. We need to make no further assumptions: by considering perfect fluids and taking the Carrollian limit, Bjorken flow arises naturally. 

\smallskip

\noindent {\em{\underline{Beyond perfect fluids}}}. Consider now leading viscous effects in Bjorken flow. EM tensor with leading-order viscous corrections in the Landau frame is
\be{eq:stress_2}
T^{\mu\nu} = \left(\epsilon + P\right) u^\mu u^\nu + P g^{\mu\nu} - \eta \sigma^{\mu\nu} - \zeta \Theta \Delta^{\mu\nu},
\ee
where $\eta, \zeta$ are shear and bulk viscosities of the fluid, $\Delta_{\mu\nu} = g_{\mu\nu} + u_\mu u_\nu$ is the projector orthogonal to the fluid velocity $u^\mu$, $\Theta = \nabla\cdot u$ is the expansion, and $\sigma^{\mu\nu}$ is the shear tensor
\begin{equation}{}
\sigma^{\mu\nu} = \frac{1}{2} \Delta^{\mu\alpha} \Delta^{\nu\beta} \left(\nabla_\alpha u_\beta + \nabla_\beta u_\alpha - \frac{2}{3} \Theta g_{\alpha\beta}\right).
\end{equation}
The viscosities $\eta,\zeta$ are non-negative and lead to dissipation. Specializing to Bjorken flow with the velocity profile $u^\mu = (1,0,0,0)$ in Milne coordinates, one has 
\be{}
\Theta = \frac{1}{\tau}\, ,\quad \sigma^{\tau\tau} = \sigma^{\tau\rho} = \sigma^{\rho\tau} = 0\, ; \quad \sigma^{\rho\rho} = \frac{2}{3\tau^3}.
\ee 
This leads to $T^{\rho\rho} = \frac{P}{\tau^2} - \frac{1}{\tau^3}\left(\frac{2\eta}{3}+\zeta\right)$, while $T^{\tau\tau} = \epsilon(\tau), T^{\tau\rho} = T^{\rho\tau} = 0$, as before. The hydrodynamic equation \eqref{eq:hydroeqn} now gives
\be{non-per-bjorken}
\frac{{\rm d}\epsilon}{{\rm d}\tau} = - \frac{\epsilon + P}{\tau} +  \frac{1}{\tau^2}\left(\frac{2\eta}{3}+\zeta\right). 
\ee
Given the equation of state $P = P(\epsilon)$, \eqref{non-per-bjorken} completely determines the time evolution of the fluid, now including the leading dissipative effects as well.

We can add in dissipation on the Carroll side as well. 
Restoring the factors of $c$, the non-perfect EM tensor \eqref{eq:stress_2} is
\begin{equation}\label{non-perfT}
    T^{\mu\nu}=(\epsilon+P)\frac{u^{\mu}u^{\nu}}{c^2}+Pg^{\mu\nu}+\tau^{\mu\nu},
\end{equation}
with $u^{\mu}\tau_{\mu\nu}=0$. Then as $c \rightarrow 0$, the non-perfect part behaves as \cite{Ciambelli:2018wre}
\begin{equation}\label{tau_behaviour}
    \tau^{ij} = - \, \Xi^{ij}+\mathcal{O}(c^2)\, .
\end{equation}
For the general fluid velocity profile \eqref{eq:gen_para}, and assumptions \eqref{eq:Carroll_Limit_1} and \eqref{tau_behaviour}, the Carrollian fluid equations are now 
\begin{subequations}\label{eq:nonperfCarroll}
\begin{align}
    \frac{1}{\Omega}\partial_t\varepsilon &=-\theta\left(\varepsilon+p-\frac{\Xi}{3}\right)+\Xi^{ij}\xi_{ij}\,,\label{non-perfCE}
    \\
    \hat{\partial}_i p &=-\varphi_i\left(\varepsilon+p\right)+\left(\hat{\nabla}_j+\varphi_j\right)\Xi^{j}_{\hphantom{j}i}\nonumber\\
	&\quad-\left(\frac{1}{\Omega} \partial_t + \theta\right) \left[(\varepsilon+p)\beta_i - \beta_j \Xi^j_{\hphantom{j}i}\right],
\end{align}
\end{subequations}
where $\Xi \equiv \Xi^{ij} a_{ij}$, $\xi_{ij} \equiv \frac{1}{2\Omega}\partial_t a_{ij}-\frac{\theta}{3}a_{ij}$ defines the Carrollian shear, and $\hat{\nabla}_i$ is the Levi-Civita-Carroll covariant derivative \footnote{This acts as $\hat{\nabla}_i V^j=\hat{\partial}_i V^j+\hat{\gamma}^j_{ik}V^k$. The Levi-Civita-Carroll connection coefficients are given via $\hat{\gamma}^{i}_{jk}=\frac{a^{il}}{2}\left(\hat{\partial}_ja_{kl} + \hat{\partial}_ka_{jl}-\hat{\partial}_la_{jk}\right).$ For more details, see \cite{Ciambelli:2018xat}.}.

For the relativistic viscous stress tensor \eqref{non-perfT}, we have up to first order in the derivative expansion 
\begin{equation}
    \tau^{\mu\nu}=-\eta\sigma^{\mu\nu}-\zeta \Theta \left(g^{\mu\nu} + \frac{u^{\mu}u^{\nu}}{c^2}\right).
\end{equation}
Taking $c\rightarrow 0$, and under the assumption \cite{Ciambelli:2018xat}
\begin{equation}
\label{eq:Carroll_Limit_2}
\eta=\Tilde{\eta}+\mathcal{O}(c^2), \quad \zeta=\Tilde{\zeta}+\mathcal{O}(c^2),
\end{equation}
we get
\begin{equation}
    \Xi_{ij}=\Tilde{\eta}\xi_{ij}+\Tilde{\zeta} \theta a_{ij}.
\end{equation}
Employing the choice of the PR gauge \eqref{milneCarroll}, the Carrollian equations \eqref{eq:nonperfCarroll} now give 
\begin{equation}
    \partial_\tau\varepsilon=-\frac{\varepsilon+p}{\tau}+\frac{1}{\tau^2}\left(\frac{2\Tilde{\eta}}{3} + \Tilde{\zeta}\right), 
 \quad   \partial_i p=0\,.
\end{equation}
This matches exactly with \refb{non-per-bjorken}, courtesy of  \eqref{eq:Carroll_Limit_1} and \eqref{eq:Carroll_Limit_2}. It is thus clear that Carroll hydrodynamics encompasses Bjorken flow, including leading viscous corrections, based purely on the symmetries of the null manifold on which the ultrarelativistic fluid flows. 

\smallskip

\noindent {\em{\underline{Discussion}}}. The key message of the present work is that Carrollian hydrodynamics, instead of being exotic and lacking in applications, as has been expected in the literature so far, is everywhere. It serves as a very good approximation to any fluid moving near the speed of light and thus is of relevance to a whole host of very fundamental physical situations, from the very early universe to the QGP. 

The importance of our rather simple observations in this paper cannot be overstated. There are numerous directions that this work has opened up. Let us briefly comment upon some of them. 

\noindent $\bullet$ {\em{An organizing principle.}} We have shown that Carroll symmetry governs the Bjorken flow, which very well approximates the central rapidity region of heavy-ion collisions and is vital to the understanding of the QGP. As we move away from the central rapidity region, corrections to this leading behaviour are expected to lead to a lot of physics that is or can be experimentally tested e.g.\ at the LHC. Our observations in this paper pave the way for a systematic expansion (in powers of $c$) to access physics beyond the central rapidity region. We have thus found an organizing principle that will be a very powerful tool going forward and will have applications in fields like heavy-ion collisions, cosmological phase transitions etc.\ to name a few. 

\noindent $\bullet$ {\em{Black holes.}} Motivated by Damour's seminal observation \cite{Damour:1978cg} that the event horizon of a black hole could be thought of as a fluid, a lot of work has been done on the so-called Membrane Paradigm \cite{Price:1986yy}. Very recently, in \cite{Donnay:2019jiz}, it was shown that membrane equations could be reinterpreted as equations of Carrollian hydrodynamics. This is again natural since the event horizon is a null surface and hence a Carrollian manifold. Carrollian symmetry on black hole horizons has also been investigated in \cite{Penna:2018gfx, Redondo-Yuste:2022czg}. Since we have shown that Carrollian hydrodynamics also accounts for Bjorken flow, there is hence an obvious relation between the Damour and Raychaudhuri equations governing the membrane paradigm and ultrarelativistic Bjorken flow. 

\noindent $\bullet$ {\em{Holography.}} The fluid-gravity correspondence relates fluids in $d$ dimensional flat spacetimes with gravity in AdS$_{d+1}$ \cite{Bhattacharyya:2007vjd}. We have shown in this work that the ultra-relativistic regime of fluids, and hence the very high energy sector, is governed by Carrollian hydrodynamics. The flat space fluid-gravity correspondence relates asymptotically flat spacetimes in $(d+1)$ dimensions with $d$-dimensional Carrollian fluids \cite{Ciambelli:2018wre}. Thus, all of the above leads to the statement that flat holography exists as a very high energy subsector of AdS/CFT. This is in keeping with various similar observations in the literature \cite{Susskind:1998vk, Polchinski:1999ry, Giddings:1999jq}. Bjorken flow has been previously addressed from the point of view of AdS/CFT (see e.g. \cite{Janik:2005zt, Janik:2006ft, Beuf:2009cx} for early works). It would be of interest to revisit these works in the light of our new perspective. 

\smallskip

We have established a novel duality between the phenomenological Bjorken flow and the geometric picture of Carrollian hydrodynamics. For a duality to prove useful, one should learn new things about both sides of the duality pair from each other. The Bjorken side has already provided the answer to the fundamental question ``What is a Carroll fluid?". Previously, Carroll fluids were thought of as mere mathematical curiosities. Our map provides a concrete example of the expectation that Carroll fluids arise whenever fluid velocities  nears $c$ {\footnote{This is at least one class of Carroll fluids, --- there may be other instances, e.g. in condensed matter situations like fractonic fluids.}}. We are currently developing a map between generic null and Carroll fluids and also constructing other explicit duality examples to further strengthen our proposal. On the other hand, the Carroll framework allows us address vital questions on the Bjorken side, including the systematic relaxation (in terms of a $c$ expansion) of the Bjorken assumptions in an effort to understand QGP away from this limit. We hope to report on these and a variety of other issues in the very near future. But it is abundantly clear that physics on both sides of the map would benefit enormously from this duality.

\bigskip
\smallskip

\noindent {\em{\underline{Acknowledgments}}}. 
We thank David Rivera-Betancour and Matthieu Vilatte for initial collaboration and Jelle Hartong, Daniel Grumiller, Shahin Sheikh-Jabbari and Marios Petropoulos for comments. 

The work of AB is partially supported by a Swarnajayanti fellowship (SB/SJF/2019-20/08) from the Science and Engineering Research Board (SERB) India, the SERB grant (CRG/2020/002035), and a visiting professorship at \'{E}cole Polytechnique Paris. AB also acknowledges the warm hospitality of the Niels Bohr Institute, Copenhagen during later stages of this work. 
The work of AS is supported by the European Research Council (ERC) under the European Union's Horizon 2020 research and innovation programme (grant agreement no.\ 758759).

\bibliography{bjorken-carroll-PRL}

\end{document}